\documentclass[twocolumn,showpacs,preprintnumbers,amsmath,amssymb,aps,prl,superscriptaddress]{revtex4-1}
\usepackage{color}
\usepackage{graphicx}
\usepackage{textcomp} 
\usepackage{subeqnarray}
\usepackage[caption=false]{subfig} 
\begin{document}
\title{Waltzing route towards double-helix formation in cholesteric shells}
\author{Alexandre  Darmon}
\affiliation{EC2M, UMR CNRS 7083 Gulliver, ESPCI, PSL Research University, 10 rue Vauquelin, 75005 Paris, France}
\author{Michael Benzaquen}
\affiliation{EC2M, UMR CNRS 7083 Gulliver, ESPCI, PSL Research University, 10 rue Vauquelin, 75005 Paris, France}
\affiliation{Current address: Capital Fund Management, 23 rue de l'Universit\'e, 75007 Paris, France}
\author{Olivier Dauchot}
\affiliation{EC2M, UMR CNRS 7083 Gulliver, ESPCI, PSL Research University, 10 rue Vauquelin, 75005 Paris, France}
\author{Teresa Lopez-Leon}
\affiliation{EC2M, UMR CNRS 7083 Gulliver, ESPCI, PSL Research University, 10 rue Vauquelin, 75005 Paris, France}

\begin{abstract}
{We study cholesteric order in liquid crystal shells with planar degenerate anchoring. We observe  that the bipolar and radial configurations intensively reported for bulk droplets have a higher degree of complexity when the liquid crystal is confined to a spherical shell.  The  bipolar configuration is replaced by a structure where the boojums are linked to a stack of disclination rings that spans the shell, while the radial configuration exhibits a double helix structure where two disclinations wind around each other. Our results confirm recent numerical simulations \cite{Sec2012} and highlight the complexity of the defect structures arising when cholesteric liquid crystals are confined to spherical geometries. We also show that the position of the boojums is only ruled by the shell geometry, independently of the cholesteric pitch. To understand quantitatively this behavior, we develop a simple yet insightful theoretical framework which captures the essence of the observed phenomenology. We also show that the transition between the two configurations is solely governed by the confinement ratio $c=h/p$, where $h$ is the average shell thickness and $p$ is the cholesteric pitch. Finally, we perform a dynamical study of this transition, and report a fascinating defect waltz due to a chemical Lehmann effect.}
\end{abstract}
\maketitle


Liquid crystal droplets are fascinating systems from both fundamental \cite{Dubois-Violette1969,Drzaic1995, Lopez-Leon2011b, Sec2014, Orlova2015,Keber2014} and applied points of view \cite{Lin2011, Bunning2000, Humar2010, Geng2013}. The curvature of the droplet surface induces geometrical frustration in the molecular arrangement of the liquid crystal, resulting in zero-order regions called defects \cite{deGennes1993, Mermin1979}. The number and nature of defects depend on i) the symmetry of the liquid crystal and ii) the type of molecular anchoring at the droplet surface. Two configurations arise when the liquid crystal has helical, or cholesteric, order and molecules are tangentially anchored to the droplet surface \cite{Xu1997}. At low values of radius to helical pitch ratios, $R/p$, cholesteric droplets are characterised by a twisted bipolar defect structure, similar to the one observed in nematics, where two surface defects or \textit{boojums} \cite{Mermin1977} appear at one diameter distance \cite{Lavrentovich1998, Lopez-Leon2011b}. Interestingly, at large values of $R/p$, the bivalent configuration is replaced by a monovalent configuration, characterised by a central bulk defect connected to a radial line-defect or disclination. This structure has drawn a lot of attention, since it shows an intriguing analogy to the Dirac monopole \cite{Lavrentovich1998}, an elementary magnetic charge. Besides, it plays a crucial role in a variety of phenomena ranging from biological morphogenesis \cite{Bouligand2011} to optical resonance \cite{Humar2010}. Despite its great interest, the exact nature of this radial defect remains uncertain \cite{Kurik1982, Bezic1992, Xu1997}. Recent numerical simulations have shown that it consists of two distinct disclinations that wind around each other in a double helix structure \cite{Sec2012}. This configuration, denoted to as radial spherical structure, RSS, was observed together with other intriguing defect structures, such as the diametric spherical structure, DSS, where a stack of disclination rings span the drop diameter. These two configurations have never been observed in experiments.

 \begin{figure}[h!]
 \centering
 \includegraphics[height=3.2cm]{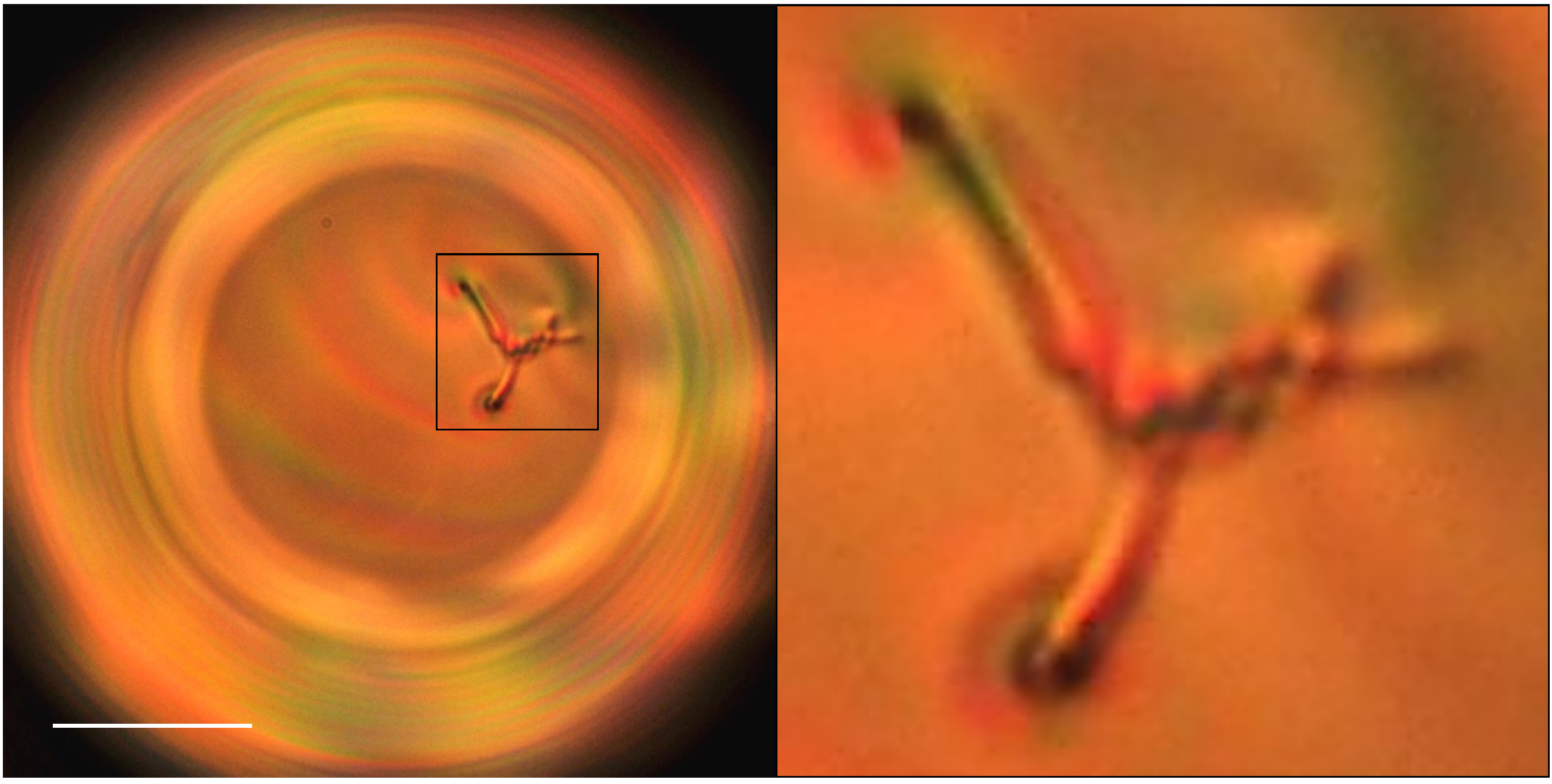}
  \vspace{-0.2cm}
 \caption{Cross-polarized image of a cholesteric shell (with $p=3.5\,\mu m$) during the formation of a double-helix structure, where two defect lines wind around each other to form a metastable braided structure. Scale bar: $20\,\mu m$.}
 \label{Torsade}
 \vspace{-0.47cm}
\end{figure}

In this paper, we take advantage of the possibility that liquid crystal shells offer in terms of controlling defect positions to study cholesteric order in spherical geometries \cite{Lopez-Leon2011, Sec2012b}. We observe that the bipolar and radial configurations, intensively reported for bulk droplets,  have a higher degree of complexity in shells. The bipolar structure is replaced by a configuration reminiscent to the DSS, where the boojums are linked to two independent stacks of disclination rings spanning the shell. We show that the equilibrium positions of the boojums in the DSS is actually governed by the geometry of the shell, independently of the cholesteric pitch. We develop a simple yet insightful theoretical model that quantitatively reproduces this behavior.  In monovalent shells, we bring experimental proof of the double helix defect structure reported in simulated droplets \cite{Sec2012}. We show that the stability of the DSS against the RSS depends on the sole confinement ratio, $c=h/p$, where $h$ is the shell thickness, and that the transition between the two configurations is first order. We finally study the dynamics of this transition; we observe that the defects of the DSS approach each other following a fascinating waltzing route, see Fig.~\ref{Torsade}, before jumping to the RSS. We relate the observed defect waltz to a chemical Lehmann effect.

\begin{figure}[t!]
 \centering
 \includegraphics[width=8.76cm]{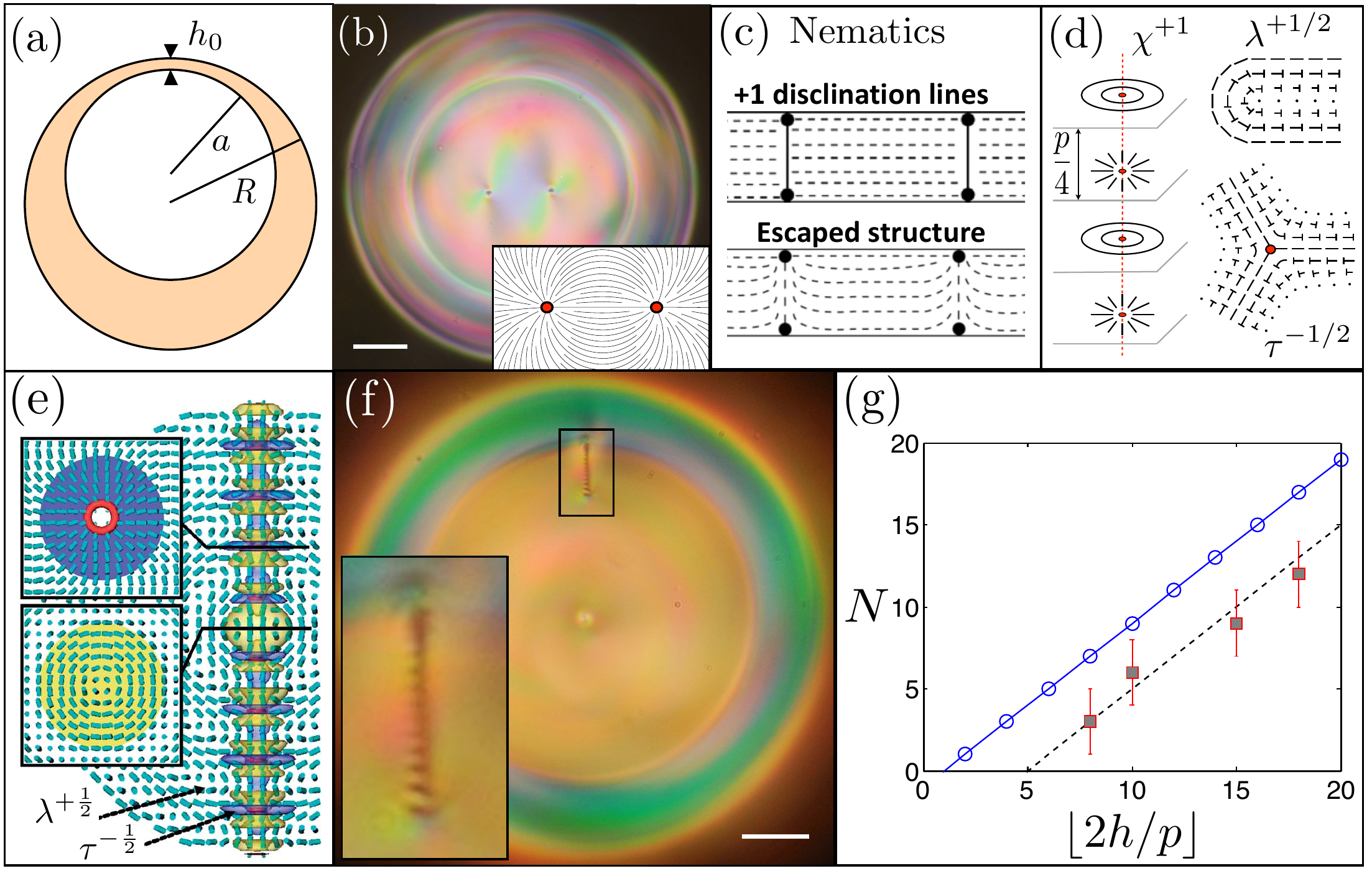}
  \vspace{-0.5cm}
 \caption{Cholesteric shells with bivalent defect structure. (a) Schematics showing the geometry of the shell. (b) Top view of a cholesteric liquid crystal shell displaying a DSS structure. The director field around either pair of surface defects is sketched in the inset. 
 (c) Planar and escaped structures of +1 disclinations in nematics. (d) $\chi^{+1}$, $\lambda^{+1/2}$ and $\tau^{-1/2}$ disclinations in cholesterics. The nail represents an out-of-plane director field, where the position of the nail indicates the extremity pointing upwardly. (e) Simulation of a diametric spherical structure (DSS) in a cholesteric droplet. (f)  DSS in a cholesteric shell.  (g) Number of disclination rings as a function of $\lfloor{2h/p}\rfloor$ in a DSS structure, from numerical predictions (blue) and observations (red). (b) and (f) are cross-polarized images, with a scale bar of $20\,\mu m$. Image (e) has been taken from \cite{Sec2012}.} 
 \label{DSS}
 \vspace{-0.5cm}
\end{figure}

We fabricate cholesteric liquid crystal shells by the means of a microfluidic device \cite{Utada2005}. Each shell is composed of an aqueous inner droplet, which is contained inside an outer liquid crystalline droplet, and in turn dispersed in an aqueous environment (see schematics in  Fig.~\ref{DSS}(a)). Typical values of the outer radii, $R$,  range between $50 \, \mu m$ and $100 \, \mu m$. Due to a density mismatch between the inner aqueous phase and the liquid crystalline phase, the inner droplet either sinks or floats inside the liquid crystal. However, the disjoining pressure prevents contact between the two droplets, resulting in a non-zero minimal thickness denoted $h_0$. Our shells are thus heterogeneous in thickness, with average thickness $h\equiv R-a$. The liquid crystalline phase is composed of a mixture of 4-Cyano-4'-pentylbiphenyl (5CB), which is nematic at room temperature, and a chiral dopant (S)-4-Cyano-4'-(2-methylbutyl)biphenyl (CB15). The cholesteric pitch $p$ is determined by the amount of dopant present in the solution. The two aqueous phases contain 1\%wt polyvinyl alcohol (PVA) to (i) stabilise the double emulsions against coalescence and (ii) ensure that the liquid crystalline molecules are parallely anchored to the two surfaces confining the shell. Under these boundary conditions, 
the molecular alignement at the spherical surfaces is frustrated, resulting in singularities in the orientational molecular order $\boldsymbol{n}$. These singularities are topological defects, which induce a $2\pi s$ rotation of $\boldsymbol{n}$, where $s$ is defined as the defect charge \cite{Mermin1979}. The topological constraints for our shells are established by the Poincar\'e theorem, which states that the total charge on each of the two spherical boundaries must be +2 \cite{Poincare1885, Hopf1926, Kamien2002}.

\begin{figure}[t!]
 \centering
 \includegraphics[height=2.4cm]{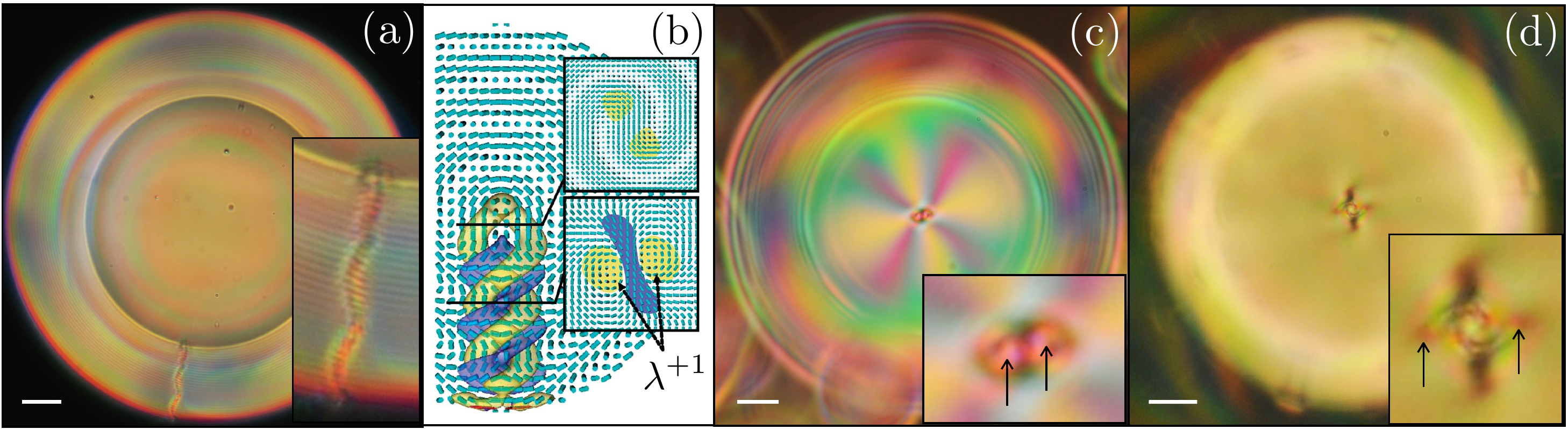}
 \vspace{-0.5cm}
 \caption{Cholesteric shells with radial spherical structure (RSS). (a), (c), (d) Experimental RSS shells with different helical pitch:  $2.7\,\mu m$, $3.5\,\mu m$, $9.3\,\mu m$, respectively. (b) Simulated cholesteric droplet with RSS. (a), (c) and (d)  are cross-polarized images, with a scale bar of $20\,\mu m$. Image (b) has been taken from \cite{Sec2012}.}
 \label{RSS}
 \vspace{-0.5cm}
\end{figure}
 
Bivalent cholesteric shells appear when chirality is low. They possess two $+1$ defects at each of the shell boundaries, such that the total charge on each sphere is $+2$, consistently with the topological requirements.  These defects appear close to each other, as imaged in Fig.~\ref{DSS}(b), which is a cross polarised picture of the top of the shell. A similar configuration appears on the inner sphere. The director field around either pair of surface defects is sketched in the inset of  Fig.~\ref{DSS}(b). Keeping this type of in-plane director field through the shell thickness would imply the formation of two $+1$ discination lines spanning the shell. 
However, 
the liquid crystalline structure is free to evolve between the boundaries to minimize its free energy. In nematic shells, the bulk elastic energy associated to $+1$ disclinations is released by allowing $\boldsymbol n$ to escape into the third dimension, as schematically represented in Fig.~\ref{DSS}(c), which  shows a local cross section of a nematic shell made by a plane containing the defects \cite{deGennes1993}. In contrast, three types of disclinations are allowed by the symmetry of cholesterics: $\chi$, $\tau$ and $\lambda$ \cite{Oswald2005}. The $\chi$ and $\tau$ lines are singular in the sense that their core correspond to a discontinuity in the order parameter, whereas $\lambda$ lines have no singular core. A diametrical structure with a $+1$ disclination spanning the entire droplet diameter was theoretically predicted for chiral droplets \cite{Bezic1992}. This structure was thought to be a $\chi^{+1}$ disclination \cite{Kurik1981}, see Fig.~\ref{DSS}(d), until recent numerical simulations suggested otherwise: the $\chi^{+1}$ line actually relaxes into an alternation of $\tau^{-1/2}$ and $\lambda^{+1/2}$ disclination rings \cite{Sec2012}, represented in the case of a droplet in Fig.~\ref{DSS}(e). The detailed structures of those defects are schematically represented in Fig.~\ref{DSS}(d). To investigate the bulk structure of the $+1$ defects observed in our cholesteric shells, we induce a slight rotation in the double emulsions by gently moving the sample, which allows us to obtain a side view of the shell, see Fig.~\ref{DSS}(f). Instead of a smooth optical texture (escaped structure) or a solid line (singular disclination), we observe an alternance of bright and dark short lines when imaging the sample between cross polarizers, see Fig.~\ref{DSS}(f). This optical texture remains unaltered after a rotation of the sample, revealing the existence of a sequence of singular bulk defect cores, reminiscent of the ring structure observed in Fig.~\ref{DSS}(e). In the droplet simulations, the number of singular rings along the radius of the droplet is  $N= \left \lfloor{2R/p}\right \rfloor -1$, where $\left \lfloor \, \right \rfloor$ denotes the integer part  \cite{Sec2012}. Applied to shells, one expects $N= \left \lfloor{2h/p}\right \rfloor -1$. We count the number of rings $N$ in four bivalent shells with different $h/p$, plot it as a function of $\left \lfloor{2h/p}\right \rfloor$ (Fig.~\ref{DSS}(g)), and find that the above relation holds, up to some offset. We attribute this offset to the fact that the lengths of the disclinations are smaller than $h$, because the defects always regroup in the thinnest part of the shell \cite{Lopez-Leon2011}.
This comforts the idea that the observed structure corresponds indeed to the numerically predicted structure  \cite{Sec2012}. This is the first experimental evidence of a DSS.

\begin{figure}[t!]
 \centering
 \includegraphics[height=5.75cm]{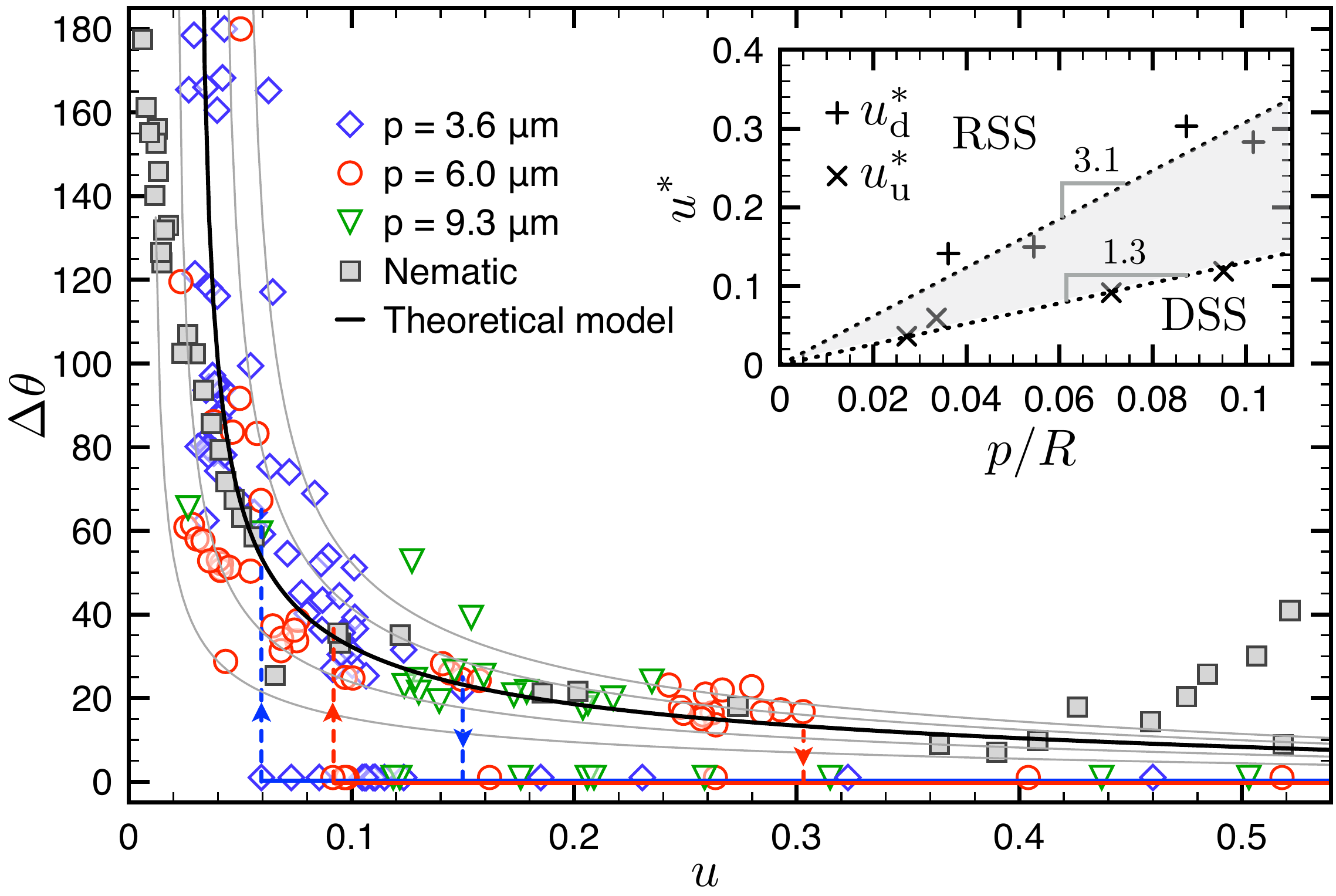}
  \vspace{-0.5cm}
 \caption{Evolution of the order parameter $\Delta\theta$ as a function of $u=h/R$ for different mixtures of 5CB and CB15, with $p$ = 9.3, 6, and 3.6\,$\mu m$. The data for $p\rightarrow \infty$ have been taken from \cite{Lopez-Leon2011}. The solid lines represent the theoretical curves, for different values of the minimal thickness $h_0$, obtained from solving $f=0$ in Eq.~\eqref{force}. Note that the phase transition cycles are only plotted for the blue and red data to facilitate reading. The inset represents the critical jumps $u^*$ as a function of $p/R$, and the grey region corresponds to the DSS/RSS coexistence zone.}
 \label{deltatheta=f(u)}
 \vspace{-0.4cm}
\end{figure}

Monovalent cholesteric shells appear at higher chirality. They display a radial defect, see Fig.~\ref{RSS}(a), in a configuration that is reminiscent to the spherulitic or Frank-Pryce texture widely reported for droplets \cite{Xu1997}. Although initially considered as $\chi^{+2}$ disclinations, the nature of the $+2$ defects appearing in this radial configuration is still an open question \cite{Kurik1982, Xu1997}. Recent numerical simulations have revealed a fascinating structure where two $\lambda^{+1}$ disclination lines wind around each other to form a double helix \cite{Sec2012}, see Fig.~\ref{RSS}(b). The ends of the $\lambda^{+1}$ lines produce a pair of close $+1$ boojums at the droplet surface, whereas they form a monopole at the center of the droplet. As of today, no experimental proof of this intricate structure, referred to as RSS, has been provided. Figure~\ref{RSS}(c), shows a top view of a mono-valent shell. The eight color brushes (pink and yellow) emerging from the center of the shell indicate a global topological charge of $+2$. However, a close view of the central defect, see the inset of Fig.~\ref{RSS}(c), reveals that it is actually composed of two $+1$ boojums, as predicted by the simulations. We find that the distance between the two boojums becomes shorter with decreasing $p$. By making this distance sufficiently large, we are able to optically distinguish two lines winding around each other, see Fig.~\ref{RSS}(d). Note that although $\lambda^{+1}$ lines are escaped, the director field is strongly distorted around it, which explains why one sees micron-thick lines \cite{Lavrentovich2001}. This the first experimental evidence of  the numerically predicted RSS in cholesterics \cite{Sec2012}.

We now investigate the phase diagram for the DSS and RSS configurations, as a function of the relevant dimensionless parameters of the problem. A cholesteric droplet is characterized by two length scales: its radius $R$, and the pitch $p$. The transition is thus triggered by the sole dimensionless ratio $R/p$ \cite{Xu1997}. In shells however, assuming that the minimal thickness $h_0$ is constant, there are three characteristic length scales, namely $h$, $R$ and $p$, from which two dimensionless parameters $h/R$ and $p/R$ can be built, if $R$ is chosen as the unit length scale.

\begin{figure*}[t!]
 \centering
\includegraphics[height=4.44cm]{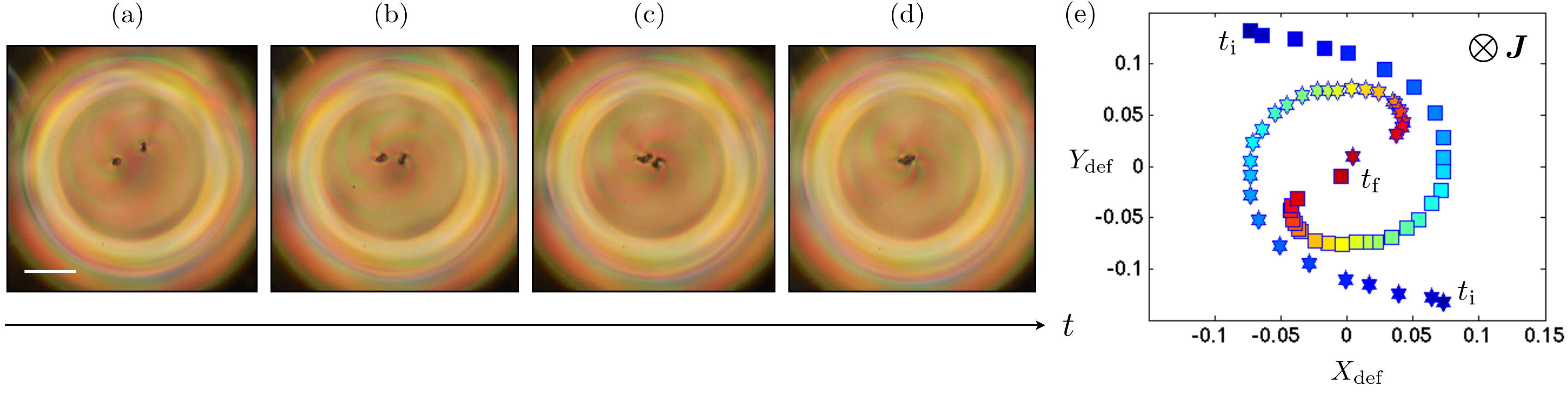}
\vspace{-0.35cm}
\caption{Dynamical transition from the DSS to the RSS. (a-d) Top view pictures of the shell taken during the transition. In this experiment, the disclinations are located at the bottom of the shell, where it is thinnest. (e) Renormalized trajectories of the outer surface defects obtained from plotting $Y_{\text{def}}=(y_{\text{def}}-y_{\text{cent}})/{R_{\text{init}}}$ versus $X_{\text{def}}=(x_{\text{def}}-x_{\text{cent}})/{R_{\text{init}}}$, where $R_{\text{init}}$ and $(x_{\text{cent}},y_{\text{cent}})$ respectively designate the initial outer radius and the $(x,y)$ coordinates of the center of mass of the trajectory. The progressive color variation indicates the temporal evolution of the system, from the beginning (blue) to the end of the deswelling experiment (red). Each symbol - star or square - is associated to one defect. (a-d) are cross-polarized images, with a scale bar of $20\,\mu m$. }
\label{trajectories}
\vspace{-0.4cm}
\end{figure*}

We first plot $\Delta\theta$, namely the angular distance between the outer surface defects, as a function of $u=h/R$ for different values of the cholesteric pitch, see Fig.~\ref{deltatheta=f(u)}. Each point on this plot refers to a different shell, at equilibrium, with its specific geometry. Positive values of $\Delta\theta$ correspond to bivalent shells, while monovalent shells correspond to $\Delta\theta=0$.  Cholesteric shells with different chirality ($p$ = 9.3, 6, and 3.6\,$\mu m$) are represented by open symbols. The bivalent nematic case, for which $p\rightarrow \infty$, is also included in the plot and represented by filled squares \cite{Lopez-Leon2011}. Starting with the bivalent shells, we observe that $\Delta\theta$ decreases when increasing $u$, first rapidly then slowly, independently of the nematic or cholesteric nature of the shells. Therefore, the defect locations depend only on the geometry of the shell, and then, are independent of the cholesteric pitch. We observe that above
 a certain critical value of $u$ denoted $u^*_\text{d}$, which depends on the pitch, no DSS are found, only RSS ($\Delta\theta=0$ in Fig.~\ref{deltatheta=f(u)}). Conversely, below a critical value for $u$ denoted $u^*_\text{u}$, smaller than $u^*_\text{d}$ and also dependent on the cholesteric pitch, we observe no RSS. The discontinuity in $\Delta\theta$, represented by descending and ascending dashed lines in Fig.~\ref{deltatheta=f(u)}, and the reported hysteresis ($u^*_\text{d}\neq u^*_\text{u}$), are a direct signature of a first order transition. 
The inset in Fig.~\ref{deltatheta=f(u)} represents the complete phase diagram for the DSS and RSS. The grey region corresponds to the ranges of parameters where both configurations coexist.
 Interestingly, the critical values $u^*_\text{u}$ and $u^*_\text{d}$ behave linearly with $p/R$. Consequently, the confinement ratio defined as $c=h/p$ is the only governing parameter for the critical jumps between DSS and RSS in cholesteric shells.  Linear regressions on the experimental jumps yield $c_\text{d}= 3.1$ and $c_\text{u}= 1.3$. 
From the above description, we highlighted the existence of a first order transition between configurations, where the angular distance between +1 defects is governed by $u=h/R$ and where the jump between configuration is controlled only by $c=h/p$.

For the bivalent configuration, the above-mentioned decreasing evolution of $\Delta\theta$, also observed in nematics both in experiments \cite{Lopez-Leon2011} and simulations \cite{Koning2013}, has been qualitatively explained in terms of two opposing forces: (i) a repulsive defect interaction of elastic nature and (ii) an attractive force due to the shell thickness gradient, which tends to bring the defects to the thinnest part of the shell. Here, we provide a quantitive expression for these forces. We start from the surface energy $E_\text{/u.l.}$ of two +1 defects interacting on a sphere of radius $R$ \cite{Nelson2002}. For small angles $\Delta\theta$, it becomes: $E_\text{/u.l.} = 2E_0 - 2\pi K \log\left({\theta}/{\sqrt 2}\right)$, where $K$ is the single elastic constant, $2\theta=\Delta\theta$, $E_0=\pi K \log(R / r_\text{c}) $ is the energy per unit length of each defects alone, and $r_\text{c}$ is the core radius \cite{Lubensky1992,Nelson2002}. We take $r_\text{c}\simeq0.01R$, corresponding to the average thickness of the DSS lines that we measured ($\simeq 1~\mu m$).
At the first level of approximation, and motivated by the fact that we are in the presence of defect lines, we write down the integrated free energy of the eccentric shell as $E = E_\text{/u.l.}  h(\theta,u)$, where $h(\theta,u)$ is the length of the defect lines, obtained from a conformal mapping technique for non-concentric shells \cite{Koning2013}. We find that $h(\theta,u) = h_0 + g(u)\,\theta^2/2 +\text{o}(\theta^2)$, where $g$ is a function of $u$ only. The force $f$ between defect lines, derived from the free energy through $f=-\partial_\theta E/R$, then reads:
\begin{eqnarray}
f=\frac{2 \pi K}{R} \left[ \frac{h_0}{\theta} + \frac{g(u)\,\theta}{2}\left(1+ 2 \log \left(\frac{\theta}{\sqrt{2}}\right) -\frac{2 E_0}{\pi K}  \right)  \right]\,.  \,\,\,\,\,\,\,\,\label{force}
\end{eqnarray}
The equilibrium position of the disclination lines is thus obtained by solving $f=0$, which yields $\Delta\theta_\text{th} = F(u,h_0)$.  Figure~\ref{deltatheta=f(u)} displays the as obtained $\Delta\theta_\text{th}$ as a function of $u$ for different values of $h_0$, ranging from 0.01R to 0.05R (solid lines). We obtain a very good agreement with the experimental data for $h_0 = 0.03R \simeq 2~\mu m$,  a reasonable value for our system. 

We take advantage of the fine control that we can achieve on the defect positions to force the transition between the DSS and RSS. 
 To perform such an experiment, we add  an aqueous solution of CaCl$_2$ to the outer phase, which imposes a difference in osmotic presure between the inner and outer aqueous phases. Since 5CB is slightly permeable to water, it acts as a membrane that allows the inner droplet to deswell. In this way, the inner radius $a$ decreases, and as a result, $u=h/R$ increases \cite{Lopez-Leon2011}. We calculate the typical viscous to elastic forces ratio, namely the Ericksen number $Er=\eta v L /K = \eta \Delta h^2 /K \Delta t$, associated to our deswelling experiment \footnote{The physical parameters to calculate the Ericksen number are: the viscosity $\eta$ of the 5CB-CB15 mixture, the typical thickness variation over the experiment $\Delta h$, the duration of the experiment $\Delta t$, and the elastic constant $K$ \cite{deGennes1993}.}.  We find that $Er \simeq 10^{-4}$, showing that our experiment can be considered as quasi-static.
The experiment starts with a DSS configuration, Fig.~\ref{trajectories}(a), where the two outer surface defects are clearly visible. As $u$ increases, the two disclinations get closer and by doing so, they turn around each other, see Fig.~\ref{trajectories}(b-c), in a fascinating \textit{waltz}, represented in Fig.~\ref{trajectories}(e) by the trajectories of the two outer surface defects. Finally, at a certain critical angular distance $\Delta\theta_{\text{crit}}$ the two lines jump together towards the center of the shell and assemble to form a double helix structure, see Fig.~\ref{trajectories}(d). In rare cases, the system can get trapped in intermediate metastable states, see Fig.~\ref{Torsade}, allowing us to shed light on the route towards of the double-helix formation.  Remarkably, performing the reverse experiment, where salt is added to the inner phase, we observe that the +2 disclination split out into two independent +1 disclinations. The rotational behavior contrasts sharply with the nematic case, in which the defects follow geodesic lines without this intriguing rotation. 
 A natural explanation for this observation is the so-called Lehmann effect \cite{Oswald2009,deGennes1993}, and more specifically the overlooked chemical version \cite{Tabe2003, Svensek2006}. In our geometry, the outward current $\boldsymbol J$ of water molecules, which diffuse through the liquid crystal, is quasi-radial and directed along the helical axis. Due to the absence of mirror symmetry of the molecules, the current induces a torque $\boldsymbol \Gamma$ on the liquid crystal molecules, resulting in a global rotation of the director field. The torque is linearly related to the current through $\boldsymbol \Gamma = -\nu \boldsymbol J$, where $\nu$ is called the Lehmann coefficient. Constitently with both swelling and deswelling experiments, we always find that $\boldsymbol \Gamma \cdot \boldsymbol J > 0$, see Fig.~\ref{trajectories}(e), such that for our right-handed cholesteric mixture $\nu<0$. 
Chirality therefore induces one main effect in the dynamics, namely the winding trajectory of the defects, clearly visible on Fig.~\ref{trajectories}(e).

We presented new defect structures in cholesteric liquid crystal shells. For the first time, experimental proofs of the numerically predicted structures of \textit{Se\u{c} et al.} \cite{Sec2012} for cholesterics in spherical geometries were highlighted. In particular, we brought evidence of a double-helix structure for twisted enough shells, and described the discontinuous transition occurring between monovalent and bivalent configurations, in both equilibrium and dynamical approaches. Our results constitute the first quantitative analysis of such intricate structures and dynamics in cholesteric shells. They open the road towards a fine characterization of the defects arrangements and dynamics in situations of practical interest \cite{Uchida2013,Uchida2015}. We thank S. \u{Z}umer, D. Se\u{c}, S. \u{C}opar for very useful discussions regarding the DSS and RSS structures, A. Fernandez-Nieves for fruitful exchanges during his last stay in Paris, invited by ESPCI ParisTech on a Paris-Sciences chair, and P. Oswald for his insight about the Lehmann rotation. We acknowledge  support from Institut Pierre-Gilles de Gennes (laboratoire d'excellence, Investissements d'avenir program ANR-10-IDEX 0001-02 PSL and ANR-10-EQPX-31), as well as the ANR with grant number 13-JS08-0006-01..

\bibliographystyle{h-physrev}
\bibliography{biblio}

\end{document}